\documentclass[a4paper]{PoS}

\title{Exclusive meson production at HERMES}

\ShortTitle{Exclusive meson production at HERMES}

\author{\speaker{Sergey Manaenkov}%
        \thanks{Thanks to the Organizing Committee for the invitation to DIS16}\\
NRC  ``Kurchatov Institute'', Petersburg Nuclear Physics Institute,\\
on behalf of the HERMES collaboration\\
       E-mail: \email{sman@mail.desy.de}}

\abstract{The data  were  accumulated with
the HERMES forward spectrometer using the 27.6 GeV longitudinally polarized electron or positron beam of
HERA.
Exclusive electroproduction of $\omega$ mesons
on unpolarized hydrogen and deuterium targets is studied in the
kinematic region of $Q^2>1.0$~GeV$^2$, 3.0 GeV $< W <$ 6.3 GeV, and
 $-t'< 0.2  $ GeV$^{2}$, while for $\rho^0$-meson production on a transversely polarized hydrogen target 
$-t'< 0.4$ GeV$^{2}$ is used.
 Spin-density matrix elements  for $\omega$ production are presented in projections of $Q^2$ or  $-t'$, while
 the ratios of the helicity amplitudes for the reaction $\gamma^*+p \to \rho^0+p$   
are obtained in the entire kinematic region.   The usage of the transversely polarized target allows for
the first time the extraction of the ratios of  certain nucleon-helicity-flip amplitudes to the natural-parity exchange amplitude 
$T_{0\frac{1}{2}0\frac{1}{2}}$ without the nucleon-helicity flip describing the longitudinal $\rho^0$-meson production by the longitudinal 
virtual photon.
Violation of $s$-channel helicity conservation   is observed  for both mesons.  
 A dominant contribution from  unnatural-parity exchange (UPE)  is established for special combinations of the $\omega$ spin-density matrix 
elements. For $\rho^0$-meson production, the UPE amplitude without the nucleon-helicity flip, $U_{1\frac{1}{2}1\frac{1}{2}}$, is shown to be the 
largest UPE amplitude. 
Good agreement is found between the HERMES data on $\omega$ production on the proton and results of the
perturbative QCD-inspired Goloskokov-Kroll model that includes pion-pole contribution.}

\FullConference{XXIV International Workshop on Deep-Inelastic Scattering and Related Subjects\\
                 11-15 April, 2016\\
                 DESY Hamburg, Germany}

\begin{document}

\section{Introduction}
Exclusive electroproduction of vector mesons ($V$) on nucleons ($N$) gives
information both on the reaction mechanisms  and the nucleon structure 
\cite{diehl1}. Electroproduction at high energies can be considered to consist of
three   subprocesses: i) the incident lepton emits a virtual photon
$\gamma^*$, which  dissociates into a  quark-antiquark pair; ii) this  $q\bar{q}$ pair
interacts strongly with the nucleon; iii) the observed vector meson is formed from the scattered $q\bar{q}$
pair. In Regge phenomenology, the interaction of the $q\bar{q}$ pair with the
nucleon proceeds through the exchanges of a pomeron or/and
exchanges of secondary regge\-ons.
If the quantum numbers of the particle lying on the Regge trajectory are
$J^P=0^+,\;1^-$, etc. (pomeron, $\rho$, $f_2$, ...), the process is denoted  Natural  Parity Exchange
(NPE). Alternatively, the case of   $J^P=0^- ,\;1^+$, etc. ($\pi$,  $a_1$, ...) corresponds to
Unnatural Parity Exchange (UPE).  In the framework of perturbative quantum chromodynamics valid at large photon virtuality $Q^2$ and high 
photon-nucleon center-of-mass (CM) energy $W$,  the nucleon structure can also be studied through hard exclusive meson production, as the process 
amplitude contains   Generalized  Parton Distributions (GPDs) (see review \cite{diehl1}). However, the factorization property  that permits to 
extract GPDs is rigorously proved in Ref.~\cite{CFS} only for the amplitude $F_{00}$ of longitudinal vector meson production by longitudinal 
virtual photons. In the Goloskokov-Kroll (GK) model (see \cite{GK} and references therein), the validity of factorization is assumed for some 
other amplitudes in addition to $F_{00}$ and this assumption is justified with a good description of the existing data. 

Recently, Spin-Density Matrix Elements (SDMEs) were studied by CLAS \cite{clas} for exclusive $\omega$ electroproduction
at 1.6 GeV$^2$ $< Q^2 < $ 5.2 GeV$^2$ and the contribution of the $\pi$-reggeon was found to be dominant even for large $Q^2$.
The presented HERMES data \cite{HER-OMEGA} can also be well described in the GK model if the pion exchange is taken into account. This means 
that the GPD-based approach should be modified at intermediate energies and $Q^2$.

All observables in vector-meson electroproduction can be expressed in terms of the helicity amplitudes in the CM system of the process 
$\gamma^* + N \to V+N$,
in particular, SDMEs are functions of the helicity amplitude ratios (HARs). Therefore HARs can be extracted from the data as was shown 
in Ref.~\cite{DC-84}. For the first time, HARs with nucleon-helicity flip are obtained in the present analysis from the 
data on $\rho^0$-meson production on a transversely polarized hydrogen target.

 \section{Helicity Amplitudes}
The angular distribution of  the  final-state particles depends on the SDMEs. In the present paper, the formalism 
proposed in Ref.~\cite{Schill} for  SDMEs $r^{\alpha}_{\lambda_V\lambda'_V}$  is used.
The  SDMEs can be expressed  in terms of the helicity amplitudes
{$F_{\lambda _{V} \lambda '_{N}\lambda _{\gamma} \lambda_{N}}$}
of the process $\gamma^{*}(\lambda _{\gamma})+N(\lambda _{N}) \to V(\lambda _{V}) +N(\lambda' _{N})$, 
where the particle helicities are given in parentheses.
The helicity amplitudes depend on $W$, $Q^{2}$, and $t'=t - t_{min}$, where
$t$ is the Mandelstam variable  and $-t_{min}$ represents the
smallest kinematically allowed value of $-t$ at fixed $W$ and $Q^{2}$. 

Any helicity amplitude can be  decomposed into a sum of a NPE ($T_{\lambda_{V} \lambda '_{N} \lambda_{\gamma}  \lambda_{N}}$) 
 and an UPE ($U_{\lambda_{V}\lambda '_{N} \lambda_{\gamma}  \lambda_{N}}$) amplitude: 
$F_{\lambda_{V} \lambda '_{N} \lambda_{\gamma}  \lambda_{N} } =
T_{\lambda_{V} \lambda '_{N} \lambda_{\gamma}  \lambda_{N} }+
U_{\lambda_{V}\lambda '_{N} \lambda_{\gamma}  \lambda_{N}}$, 
for details see Refs.~\cite{Schill,DC-24}.
The amplitudes obey the symmetry relations
that hold because of parity conservation  (see, e.g., Ref.~\cite{Schill})
\begin{eqnarray} 
T_{\lambda_{V} \lambda '_{N}\lambda_{\gamma} \lambda_{N}} &  =  &
(-1)^{-\lambda_{V}+\lambda_{\gamma}}\,
T_{-\lambda_{V} \lambda '_{N}-\lambda_{\gamma} \lambda_{N}}=(-1)^{-\lambda_{N}+\lambda'_{N}}\,
T_{\lambda_{V} -\lambda '_{N}\lambda_{\gamma} -\lambda_{N}},
 \label{symtn} \\ 
U_{\lambda_{V} \lambda '_{N}\lambda_{\gamma} \lambda_{N}}  & =   &
-(-1)^{-\lambda_{V}+\lambda_{\gamma}}\,
U_{-\lambda_{V} \lambda '_{N}-\lambda_{\gamma} \lambda_{N}}
=-(-1)^{-\lambda_{N}+\lambda'_{N}}\,U_{\lambda_{V} -\lambda '_{N}\lambda_{\gamma} -\lambda_{N}}.
  \label{symun}
\end{eqnarray} 
These symmetry properties of the helicity amplitudes permit to introduce the abbreviated notations:
\begin{eqnarray}
T^{(1)}_{\lambda_{V} \lambda_{\gamma} }  \equiv  T_{\lambda_{V} \frac{1}{2}\lambda_{\gamma} \frac{1}{2}},   \;\;
U^{(1)}_{\lambda_{V} \lambda_{\gamma} }  \equiv  U_{\lambda_{V} \frac{1}{2}\lambda_{\gamma} \frac{1}{2}}, \;\;
T^{(2)}_{\lambda_{V} \lambda_{\gamma} }  \equiv  T_{\lambda_{V} \frac{1}{2} \lambda_{\gamma} -\frac{1}{2}},   \;\;
U^{(2)}_{\lambda_{V} \lambda_{\gamma} } & \equiv & U_{\lambda_{V} \frac{1}{2}\lambda_{\gamma} -\frac{1}{2}}.
  \label{diag-non}   
\end{eqnarray}
All other amplitudes can be obtained from the symmetry relations 
given by Eqs.~$(\ref{symtn}-\ref{symun})$. The HARs extracted in the present analysis are defined for $n=1,\;2$ as
$t^{(n)}_{\lambda_{V} \lambda_{\gamma} }=\frac{T^{(n)}_{\lambda_{V} \lambda_{\gamma} }}{T^{(1)}_{00}},\;\;
u^{(n)}_{\lambda_{V} \lambda_{\gamma} }=\frac{U^{(n)}_{\lambda_{V} \lambda_{\gamma} }}{T^{(1)}_{00}}.$

\section {Data Analysis}

The data were accumulated with the HERMES spectrometer using the 27.6 GeV longitudinally polarized electron 
or positron beam of HERA, and gaseous hydrogen or deuterium targets.
The HERMES forward spectrometer is described  in Ref.~\cite{identif}.
 The spectrometer permitted a precise measurement of charged-particle momenta with a resolution of $1.5\%$. 
A separation of leptons was achieved with an average efficiency of $98\%$ and a hadron
contamination below $1\%$.

The $\omega$ and $\rho^0$  mesons are produced and decay in the following exclusive reactions:
$e + p \to e + p + V$, $\omega \to \pi^+ + \pi^- + \pi^0,\;\;\;\pi^0 \to 2 \gamma$ for $V=\omega $
and $\rho^0 \to \pi^+ + \pi^- $ when $V=\rho^0$.
The following requirements were applied for event  selection in $\omega$-meson production:\\
 i) Exactly two oppositely charged hadrons and one lepton with the same charge as the beam lepton are identified.\\
ii) A $\pi^{0}$ meson that is reconstructed from two calorimeter
clusters  is selected requiring the two-photon invariant mass to be in the interval $0.11$~GeV $ < M({\gamma\gamma})<$ 0.16 GeV.\\
iii) The three-pion invariant mass is required to obey 0.71 GeV$\le$ M($\pi^+ \pi^- \pi^0$)$\le$ 0.87 GeV.\\
iv)  The scattered-lepton momentum lies above $3.5$~GeV.\\
v) Taking into account the spectrometer resolution, the  missing energy $ \Delta E $ has to lie in the interval  -1.0 GeV $<\Delta E <$ 0.8 GeV. 
Here, $\Delta E = \frac{ M^{2}_{X} -M^{2}_{p}}{2 M_{p}}$, with $M_{p}$ being
the proton mass and $ M^{2}_{X}=({p} + {q}- {p}_{\pi^+}$ - ${p}_{\pi^-}$ -
${p}_{\pi^0})^{2}$  the missing mass squared, where  ${p}$, ${q}$, ${p}_{\pi^+}$, $
{p}_{\pi^-}$, and  ${p}_{\pi^0}$ are the four-momenta of target nucleon, virtual photon, and each of the three
 pions respectively. \\
vi) The  constraints
$ Q^2> $ 1.0 GeV$^2$, 6.3~GeV~$>W > 3$~GeV, $-t' <$ 0.2 GeV$^{2}$   are applied. 

\begin{figure}
\begin{center}
\includegraphics[height=3.5cm]{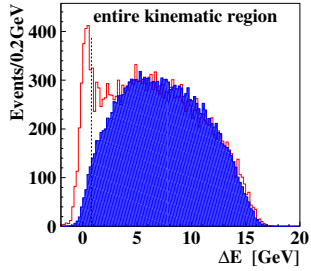}
\end{center}
\caption{ \label{deltae} 
The $\Delta E$  distribution of  $\omega$ mesons produced
in the  entire kinematic region.
The vertical dashed line denotes the upper limit of the exclusive region.}
\end{figure}

\indent After application of all these constraints, the proton sample contains 2260
and the deuteron  sample 1332 events of exclusively produced  $\omega$ mesons.
These data samples are referred to in the following  as data in the ``entire
kinematic region".  The distribution of missing energy $\Delta$E, 
shown in Fig.~\ref{deltae}, exhibits a clearly visible exclusive peak. The  shaded histogram
represents semi-inclusive deep-inelastic scattering (SIDIS)
background obtained from a PYTHIA Monte Carlo  simulation that is normalized to the data
in the region of 2 GeV $< \Delta E < $ 20 GeV.
The simulation is  used to determine the fraction of background under the exclusive peak.
It amounts to  about 20$\%$ for the entire kinematic region.

The only requirements changed for event  selection in $\rho^0$-meson production were:\\
ii)  No calorimeter clusters were observed (no $\pi^{0}$ is detected). 
iii) The two-pion invariant mass is required to obey 0.7 GeV$\le$ M($\pi^+ \pi^- $)$\le$ 1.0 GeV.
vi) The  constraint $-t' <$ 0.4 GeV$^{2}$   is applied.
The total number of reconstructed events with $\rho^0$ mesons exclusively produced on the transversely polarized proton (transverse polarization 
$P_T=0.72 \pm 0.06$) was 8741.

\section{Results}
In exclusive electroproduction, SDMEs are fitted as parameters of the experimental angular distribution
of the final particles. The  ``unpolarized'' SDMEs can be obtained from scattering of unpolarized initial particles
while the ``polarized'' SDMEs can be extracted only  from data collected  with a longitudinally polarized  beam.
The results on SDMEs are obtained using an unbinned maximum likelihood method. All details of the fit procedure 
including the method of taking into account the background corrections are described in Ref.~\cite{HER-OMEGA}.

\begin{figure}
\begin{center}
\includegraphics[height=8.5cm]
{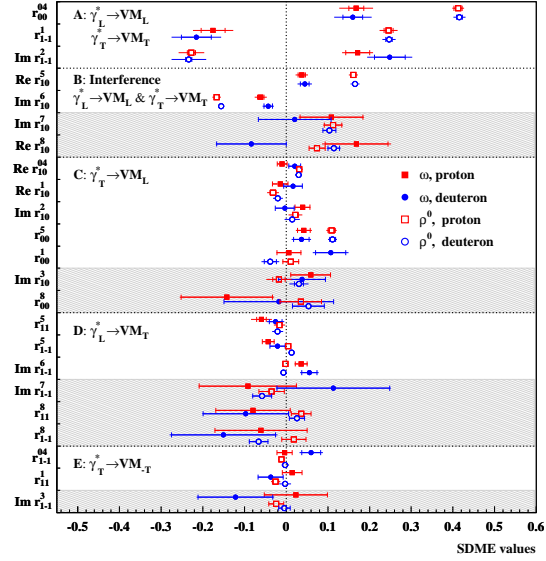}
\end{center}
\caption{\label{sdmescaled} Comparison of 23 SDMEs for $\omega$  and
$\rho^{0}$~\cite{DC-24} for the  entire kinematic region.
The inner error bars represent the statistical uncertainties, while the
outer ones indicate the statistical and systematic uncertainties added in
quadrature. Unpolarized (polarized) SDMEs are displayed in the unshaded
(shaded) areas.}
\end{figure}

\indent The  SDMEs of the $\omega$ and $\rho^0$ mesons  for the entire  kinematic region
are compared in Fig.~\ref{sdmescaled}. The SDMEs for $\omega$ are extracted
at $ \langle Q^2 \rangle = 2.42$ GeV$^2$, $\langle W\rangle =4.8$
GeV, $ \langle -t'\rangle = 0.080$ GeV$^2$ while for $\rho^{0}$ production
$\langle Q^2  \rangle = 1.95$~GeV$^2$, $\langle W\rangle = 4.8$~GeV, and  $\langle -t' \rangle = 0.13$~GeV$^2$.
These SDMEs are divided into five classes corresponding to various
helicity transitions $\lambda_{\gamma} \to \lambda_V$.
The main terms in the expressions of class-A SDMEs correspond to the transitions from
longitudinal virtual photons to  longitudinal vector mesons, $\gamma^*_L \to V_L$,  
and from transverse virtual photons to transverse vector mesons, $\gamma^*_T \to V_T$.
The dominant terms of class B correspond to the interference of these two transitions.
The main terms of class-C, D, and  E SDMEs are proportional to small amplitudes 
of $\gamma^*_T \to V_L$, $\gamma^*_L \to V_T$, and
$\gamma^*_T \to V_{-T}$ transitions respectively.\\
\indent The SDMEs for the proton and deuteron data are
found to be  consistent with one another within their combined total uncertainties.
 In Fig.~\ref{sdmescaled}, the uncertainties of the  polarized SDMEs are larger
than those of the unpolarized  SDMEs because the mean polarization $<|P_{b}|> \approx 40\%$ and 
the polarized  SDMEs are multiplied by the small factor $|P_{b}|\sqrt{1- \epsilon}\approx 0.2 $
in the equation for the  angular distribution. Here, $\epsilon$ is the ratio of fluxes of 
the longitudinally and transversely polarized virtual photons.
 The linear combination of the class-A and class-B SDMEs, which are to be zero if $S$-Channel Helicity Conservation (SCHC) approximation is 
valid, are really zero within expe\-ri\-mental uncertainties (see Ref.~\cite{HER-OMEGA}).
All SDMEs of classes C, D, E are to be zero in the SCHC approximation.
The SDME $r^{5}_{0}$ violates SCHC both for the $\rho^0$ and $\omega$ mesons since it is nonzero as is seen from Fig.~\ref{sdmescaled}.
\begin{figure}
\begin{center}
\includegraphics[height=4.25cm]{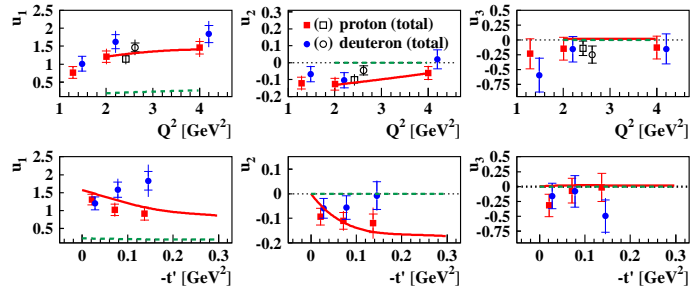}
\end{center}
\caption{ \label{u1} The $Q^{2}$ and $-t'$ dependences of $u_1$, $u_2$, and  $u_3$,
where $u_1=1-r^{04}_{00}+2r^{04}_{1-1}-2r^{1}_{11}-2r^{1}_{1-1}$, $u_2=r^{5}_{11}+r^{5}_{1-1},$ and $u_3=r^{8}_{11}+r^{8}_{1-1}.$
 The open symbols represent the values over the entire
  kinematic region.  Solid (dashed) curves are obtained in the GK model when
the pion exchange is (not) taken into account. The error bars are the same as in Fig.~2.
}
\end{figure}

As is also seen the SDMEs,  $r^{1}_{1-1}$ and
$\mathrm{Im} \{r^{2}_{1-1}\} $ have opposite sign for  $\omega$ and
$\rho^{0}$ mesons. The explanation  follows from the expressions of these SDMEs in terms of the amplitudes given in 
\cite{Schill,DC-24}: $|T_{11}|^2 > |U_{11}|^2$  for the $\rho^{0}$ meson, while $|U_{11}|^2 > |T_{11}|^2$
for the $\omega$ meson. Moreover, the contribution  of the UPE amplitude $U_{11}$ dominates in the cross section of the $\omega$-meson production
 while its contribution for the $\rho^0$-meson electroproduction is small \cite{DC-24}. As the simplest consequence of UPE 
dominance, the values of the SDME combinations $u_1$, $u_2$, and $u_3$ are unexpectedly large for the $\omega$ meson as is seen from 
Fig.~\ref{u1}. These combinations are well described in the GK model \cite{GK}, if the pion 
exchange is included in the calculation, as can be seen from a comparison of solid and dashed curves with the HERMES data. 

\begin{figure*}\centering
\includegraphics[height=9.5cm]{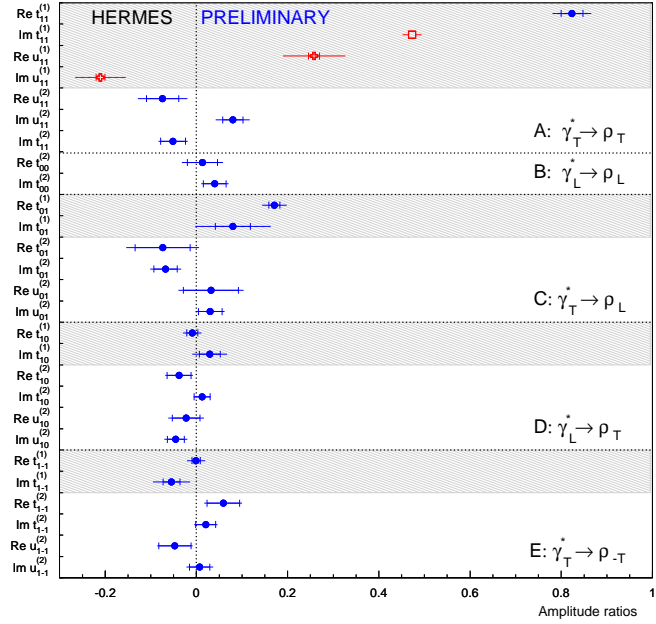}
\caption{\small Helicity-amplitude ratios in the entire kinematic region.
The inner error bars represent the statistical  uncertainty, while the outer
ones represent statistical and systematic uncertainties added in quadrature.}
\label{rel-ent-bin-25}
\end{figure*}

The HARs obtained from the 25-parameter fit in the entire kinematic region
($\langle W \rangle=4.73$ GeV, $\langle Q^2\rangle=1.93$ GeV$^2$, $\langle-t'\rangle=0.132$ GeV$^2$) are shown in 
Fig.~\ref{rel-ent-bin-25}.
While the phase of $u^{(1)}_{11}$ is fixed according to the results of
Refs.~\cite{HERL1,HERL2}, its modulus is fit so that the two crosses represent the results of fitting one free parameter.
The value of $\rm{Im}[t^{(1)}_{11}]$ (open square) represents the result of Ref.~\cite{DC-84}; its error bar
shows the total uncertainty. The shadowed area corresponds to results that are 
also obtained in Ref.~\cite{DC-84}, while all other points are calculated for the first time. As is seen from Fig.~\ref{rel-ent-bin-25},
all HARs except $t^{(1)}_{11}$, $t^{(1)}_{01}$, and $u^{(1)}_{11}$, are compatible with zero within their total uncertainties.
The HAR $t^{(1)}_{01}$ is responsible for SCHC violation, while $u^{(1)}_{11}$ shows the role of UPE in  $\rho^0$ production.

\end{document}